\documentclass[sigconf]{acmart}
\AtBeginDocument{%
  \providecommand\BibTeX{{%
    \normalfont B\kern-0.5em{\scshape i\kern-0.25em b}\kern-0.8em\TeX}}}

\setcopyright{acmcopyright}
\copyrightyear{2021}
\acmYear{2021}
\acmConference[CHI 2021]{CHI 2021: ACM CHI Conference on Human Factors in Computing Systems}{May 08--13, 2021}{Yokohama, Japan}
\acmBooktitle{CHI 2021: ACM CHI Conference on Human Factors in Computing Systems,
  May 08--13, 2021, Yokohama, Japan}
\acmPrice{15.00}

\usepackage{url}
\begin{document}

\title{Esports Agents with a Theory of Mind: Towards Better Engagement, Education, and Engineering}


\author{Murtuza N. Shergadwala}
\email{mshergad@ucsc.edu}
\affiliation{%
  \institution{University of California, Santa Cruz}
  \streetaddress{1156 High Street}
  \city{Santa Cruz}
  \state{California}
  \postcode{95064}
  \country{USA}
}

\author{Magy Seif El-Nasr}
\email{mseifeln@ucsc.edu}
\affiliation{%
  \institution{University of California, Santa Cruz}
  \streetaddress{1156 High Street}
  \city{Santa Cruz}
  \state{California}
  \postcode{95064}
  \country{USA}
}


\begin{abstract}

The role of AI in esports is shifting from leveraging games as a testbed for improving AI algorithms to addressing the needs of the esports players such as enhancing their gaming experience, esports skills, and providing coaching. For AI to be able to effectively address such needs in esports, AI agents require a theory of mind, that is, the ability to infer players' tactics and intents. To that end, in this position paper, we argue for human-in-the-loop approaches for the discovery and computational embedding of the theory of mind within behavioral models of esports players. We discuss that such approaches can be enabled by player-centric investigations on situated cognition that will expand our understanding of the cognitive and other unobservable factors that influence esports players' behaviors. We conclude by discussing the implications of such a research direction in esports as well as broader implications in engineering design and design education.

\end{abstract}





\maketitle

\section{ AI in esports}
%With the constant evolution of online gaming, electronic sports or esports lacks a precise definition. However, esports typically refers to competitive multiplayer gaming with spectators~\cite{freeman2017esports} and this paper adopts such a view of esports.

The use of AI technologies to enhance gaming experience of esports players is booming~\cite{risi2020chess}. Such technologies span player-centric use cases such as enabling players to automatically import other expert player's team compositions\cite{senpai}, providing them their game play statistics, and strategy suggestions to maximize their learning potential~\cite{omnicoach}. AI agents are also being used as decision-support tools to generate team compositions to enable players to practice with different teams as well as predict success metrics and win probabilities~\cite{hodge2017win,hodge2019win}. 

Although current AI technologies have advanced to provide game analytics~\cite{el2016game} to esports players, the player-centric needs in esports go beyond suggesting heuristics and enabling team compositions~\cite{weiss2013virtual}. Such needs include gaining skills such as understanding players' gameplay strategies, feeling a degree of sharedness in strategy with other team members, being able to infer opposing team's strategic plans, and to identify and act on opportunities within diverse situational contexts. Currently, players hope to satisfy such needs by gaining expertise through practice with human competitors not AI agents~\cite{freeman2017esports}. Freeman and Wohn~\cite{freeman2017esports} discuss that the desire to play with human players by esports players stems from the technical concern of AI agents lacking versatility. Such versatility includes the humanistic ability to interpret others’ intentions, beliefs, knowledge, and mental states from observable data which is referred to as the \textit{theory of mind}~\cite{premack1978does}.

In the context of esports environments, typically characterized as multi-agent environments, the theory of mind is routinely leveraged by human players to reason about other players who may be teammates or competitors. Understanding others' intents and plan can be crucial to both cooperate and compete at the same time. In order for AI agents to facilitate such understanding, they too would require the theory of mind. Thus, developing computational esports agents with a theory of mind has not only the potential to alleviate the current concerns by esports players about AI agents~\cite{freeman2017esports} but also has several benefits such as improving player engagement, performance, and collaboration skills~\cite{bard2020hanabi}. Moreover, esports agents with a theory of mind act as testbeds for broader impact in intelligent tutoring for education~\cite{millis2017impact}, discovery of human expertise in collaboration~\cite{pluss2019esports} which in turn would influence collaboration in engineering design teams~\cite{larsson2003making}, and development of theories for human-AI collaboration~\cite{wang2020human}. 
    
 Endowing esports agents with a theory of mind is challenging due to several reasons. First, it is challenging to make inferences on unobservable factors such as intents and beliefs of others~\cite{baker2009action}. Due to the inherent unobservability of such aspects, there is a lack of quantitative data as well as approaches to generate such data about player intent, beliefs, and desires. Second, computational models of the theory of mind~\cite{baker2011bayesian,yoshida2008game} are in their infancy and cannot be directly applied in the context of esports due to the sheer complexity of esports environments. This includes players making inferences about multiple unobservable aspects such as individual-level beliefs and intents, team-level beliefs and intents, as well as situational inferences for opportunistic planning. Third, there is a lack of esports player-centric investigations that can shed light on the mental processes of players that implicitly leverage such unobservable factors to influence players' gameplay actions~\cite{pedraza2020setting}. Thus, there is a need to investigate cognitive and situational factors that influence esports players' gameplay behaviors towards understanding their theory of mind. 
 
 In this position paper, we argue for human-in-the-loop (HITL) approaches for player-centric investigations towards developing computational models of esports agents with a theory of mind. We expand on this position statement such that we discuss (1) the current approaches for developing computational models of esports player behaviors and the research gaps towards understanding esports players' situated cognition, (2) the need for HITL approaches that integrate qualitative and quantitative inputs into the computational process of modeling intent, strategies, and tactics as a graphical probabilistic model, and (3) the implications of such approaches on esports and other broader implications.

% Research Gaps

% \begin{itemize}
%     \item Lack of explainability of agent behaviors. Required to develop trust, reliance, and collaboration with an agent as a team member.
%     \item Lack of inference of player goals and strategies. Required for shared cognition and mental model for coordination at team level with expert players. Required to tutor novice players by describing their own strategies and actions from an agent perspective.
% \end{itemize}

\section{Current approaches to modeling esports player behavior}

Player modeling, also termed as opponent modeling in some contexts~\cite{van2005opponent}, refers to the abstraction of a player's state, characteristics, and behaviors in the game~\cite{machado2011player,van2005opponent}. There also exists a taxonomy for the characteristics of the player that are modeled which includes a player's knowledge, position, strategy, and satisfaction about the game~\cite{machado2011player,machado2011player2}. The emphasis of modeling such player characteristics using AI agents has primarily been on generating believable behaviors~\cite{afonso2008agents} towards enabling human-AI interaction, improving player engagement, and customizing a game to address player needs~\cite{furnkranz2007recent,bakkes2012player,risi2020chess}. However, there is a lack of focus on player models that abstract player cognition and mental processes.

Approaches to player modeling, broadly fall under model-based, model-free, and hybrid approaches~\cite{yannakakis2013player}. Model-based approaches leverage theory mostly pertaining to human emotions for generating believable behaviors~\cite{yannakakis2015experience}. Model-free approaches assume some unknown functional mapping and leverage machine learning approaches such as Reinforcement Learning (RL) and Neural Networks to mimic player behaviors using data. Hybrid approaches leverage a combination of theory and data to model player behaviors~\cite{machado2011player}. Recently, researchers extended RL models to represent dynamic decision-making processes within complex stochastic environments~\cite{jaderberg2019human}. Further, OpenAI made great strides in beating DotA2 human players using RL~\cite{berner2019dota}. Such AI agents have been built using hybrid approaches and have been able to win games against human players. However, esports players' needs go beyond just winning the game to needs such as wanting to improve their gameplay skills~\cite{freeman2017esports}.

Improvement in esports skills requires coaching in the form of detection and explanation of strategies from observation of players' gameplay. While current advancements in player modeling can lead to complex and robust agents, they are limited in how they can be used to detect and explain tactics and intent from observed real-world human behaviors~\cite{nguyen2014strategy}. This is because such models lack representation of unobservable factors that influence player behavior such as their cognition and prior knowledge. Model-free approaches for player behavior do account for unobservable factors through latent parameter modeling~\cite{wang2018predictive}, such parameters are typically arbitrary and lack physical significance to the actual gameplay to derive meaningful explanations of player intent or strategies. Such abilities are the core competencies of the theory of mind, which is crucial for esports players to collaborate with team members as well as develop counter strategies to defeat opposing teams. Moreover, such detection and explanation of strategies is essential for esports coaching.

%Existing literature is vast on the topics of plan recognition~\cite{geib2009probabilistic,sukthankar2014plan}, activity recognition~\cite{lara2012survey,su2014activity}, and multi-agent collaboration~\cite{ccelikok2019interactive}. In such works, data is leveraged to train algorithms to recognize specific patterns of activities and enable prediction of stochastic actions of agents. However, much of this work cannot detect tactics or intent, the core competencies of theory of mind, which is crucial for esports players to collaborate with team members as well as develop counter strategies to defeat opposing teams.

Existing literature in computational cognitive psychology has demonstrated the benefits of computationally embedding the theory of mind to address topics such as plan recognition~\cite{baker2014modeling}, belief modeling~\cite{baker2012bayesian}, and intent recognition~\cite{sukthankar2014plan}. Such works demonstrate the ability to perform mental simulations of others' situation-dependent activities. While such works have made progress towards enabling a computational approach to the theory of mind, its application within esports is limited.

Computationally embedding the theory of mind in esports agent is still challenging due to several research gaps. First, there is a lack of player-centric investigations on the unobservable factors that players leverage while making decisions in esports gameplay. Such investigation will need to adopt HITL approaches to probe the mental processes of esports players. Second, in the context of esports, understanding intent and strategies is further complicated as players replan, behavior correct, and adopt opportunistic behaviors. Third, situational variations in esports environments are plenty and there is a lack of framework that enables capturing such variations in esports. To address such gaps, in the following section we argue for HITL approaches and elaborate how it would enable embedding theory of mind in esports agents.  

\section{Towards a theory of mind in Esports: The case for human-in-the-loop approaches}

HITL approaches imply leveraging cognitive power of humans to improve computational models~\cite{nunes2015survey}. Such approaches are crucial for complex situations where human expertise is required to make inferences~\cite{holzinger2016interactive}. HITL approaches enable development of socially intelligent agents, that is, agents with theory of mind capabilities ~\cite{dautenhahn1998art}. HITL is also foundational in areas such as Interactive Machine Learning~\cite{fails2003interactive} to improve algorithmic predictive accuracy. Another area of relevance is open-learner models~\cite{dimitrova2003style} where learners are the HITL who interact with computational models towards improving their understanding, involvement, and interaction with the learning environment. We assert that such HITL approaches can also be leveraged within esports to develop AI agents with a theory of mind that is capable of inferring intents and strategies of players and enabling esports coaching.

 %We argue that such criteria for leveraging HITL approaches are valid for esports where player expertise and multi-agent interaction needs to be inferred by players and explicitly encoded in computational models towards achieving human-AI collaboration or to leverage AI agents as coaches.
 
In the context of esports, we envision players would integrate qualitative and quantitative inputs into the computational models of esports players as a graphical probabilistic model. We argue that HITL approaches will enable players to interactively correct the computational model’s probabilities as well as nodes that make up the graphical model. Thus, such approaches would enable players to computationally embed their theory of mind capabilities. Currently, however, there is lack of frameworks, methods, and infrastructure to enable players-in-the-loop to encapsulate their thoughts computationally.

In order to facilitate HITL approaches in esports, we require player-centric investigations on cognitive factors and mechanisms that influence players' gameplay. Existing literature in esports recognizes the need for such investigations as well~\cite{gray2017game,pedraza2020setting}. Player-centric cognitive investigations would enable the development of a cognitive taxonomy of esports players. Such a taxonomy would provide different classes of cognitive variables and situational contexts that can be leveraged as blueprints to create a computational infrastructure for efficiently enabling HITL approaches to infer and embed cognitive and unobservable factors. In other words, player-centric investigations of cognitive mechanisms of their gameplay would provide a structure for HITL approaches to efficiently embed player's theory of mind in AI agents. 

For cognitive investigations in esports, we argue for a situated cognition perspective. Situated cognition posits that knowing is inseparable from doing~\cite{brown1989situated} such that knowledge resides within the situations and activities bound to social, physical and cultural contexts. In the context of esports, such a view is relevant because leveraging the players to make inferences about mental states would deepen our understanding of how such players infer intent and tactics as well as improve their gameplay skills by encountering variations in gameplay situation. 

%Moreover, investigating situated cognition in esports would enable us to understand how players infer intent and tactics given various situational contexts.

Our HITL approach for cognitive investigations of esports players is to leverage players to label mental states of observed behaviors using Stratmapper~\cite{ahmad2019modeling}- our visualization system designed to visualize user actions, such as moving the mouse on an interface or map. Further, StratMapper allows us to filter events and use freeform text to label behaviors indicating the time and spatial contexts. Using the generated labels, we will apply a qualitative content analysis method to categorize various cognitive and situational factors such as player attention and intent. Moreover, expert players will be interviewed about their labeling choices to further understand the cognitive mechanisms. We note that we do not prescribe our approach for investigating cognitive mechanisms of player behaviors. Rather we emphasize the need for understanding cognitive mechanisms in esports by developing HITL approaches in esports.

We emphasize that player-centric investigations of cognitive mechanisms acts as a stepping stone towards enabling HITL approaches for embedding theory of mind in computational agents. Further challenges remain such as developing the computational infrastructure to efficiently capture the intricacies of situational contexts and behavioral variations in esports, addressing scalability to enable a player community to contribute to the computational models, and developing validation approaches to evaluate the impact of such models on esports player outcomes.

\section{Implications of Embedding theory of mind in Esports}

In this position paper, we highlight the need to embed theory of mind capabilities of players within AI models to enable reasoning about players or other AI agents in the game environment. We argue for HITL approaches that combine computation and cognition such that the computational representation of mental states is reflective of player cognition and not an arbitrary representation of latent variables. We discuss the implications of such research both within the context of esports and the broader implications in serious games for education and engineering design in the context of human-AI collaboration.

\subsection{Within Esports}
Endowing esports agents with a theory of mind will result in a paradigm-shift in the way esports player interact with AI agents which could manifest in novel innovative gameplay experiences. Beyond player-agent interaction, AI agents would act as intelligent tutoring systems for esports that would improve player expertise and skills in a personalized manner beyond simple heuristics and other player specific strategies. Such AI agents would improve believability of the gameplay experience, thereby, impacting game narratives and creative story telling as well.

% \begin{itemize}
%     \item intelligent tutoring system for esports
%     \item improved player experience and engagement
%     \item improved believability and impact on game narratives and creativity
% \end{itemize}

\subsection{Broader Implications: Engineering Design and Design Education}
While the implications of AI agents with a theory of mind on human-computer interaction are plenty, we focus on the impact of esports research on engineering design and design education. We highlight the similarity of spatio-temporal interaction in esports with engineering applications such as Computer Aided Design (CAD) as well as the similarity of esports player behaviors to engineering designers. 

In esports, players collaborate in teams, make sequential and dynamic decisions, interact with their spatial environment, and their strategies evolve temporally as the gameplay proceeds. Similarly, engineering design teams collaboratively make domain-specific decisions~\cite{hazelrigg1998framework} that involve interacting with a design environment and their strategies evolve as they move from conceptual design stage to the later stages of an engineering design process~\cite{simon1988science}. We note that domain knowledge and problem-specific aspects of the engineering design process influence design outcomes~\cite{shergadwala2018quantifying,shergadwala2016understanding}. Such aspects are difficult to teach as a part of the engineering education curricula~\cite{shergadwala2018students}. The understanding of the design of AI agents with a theory of mind that would assist esports players in discovering strategies could also be leveraged by engineering design teams. Such teams would collaborate better by understanding individual team member's approach towards problem solving facilitated by the AI agent. Moreover, such agents could act as intelligent tutors for engineering design students to enable them explicitly understand the impact of their procedural decision making and problem solving strategies on design outcomes. 
% \begin{itemize}
%     \item spatiotemporal interaction similarity of games with engineering applications such as CAD 
%     \item esports involves sequential decision making which is abundant in engineering design as well...AI agents built in esports context can be leveraged to improve engineering design process
%     \item esports involves players collaborating in teams which is also a characteristic of engineering design....application in human AI intercation and collaboration to boost engineering innovation
%     \item improved believability and impact on game narratives and creativity
% \end{itemize}

\bibliographystyle{ACM-Reference-Format}
\bibliography{sample-base}


\begin{thebibliography}{46}

%%% the Web address.

\ifx \showCODEN    \undefined \def \showCODEN     #1{\unskip}     \fi
\ifx \showDOI      \undefined \def \showDOI       #1{#1}\fi
\ifx \showISBNx    \undefined \def \showISBNx     #1{\unskip}     \fi
\ifx \showISBNxiii \undefined \def \showISBNxiii  #1{\unskip}     \fi
\ifx \showISSN     \undefined \def \showISSN      #1{\unskip}     \fi
\ifx \showLCCN     \undefined \def \showLCCN      #1{\unskip}     \fi
\ifx \shownote     \undefined \def \shownote      #1{#1}          \fi
\ifx \showarticletitle \undefined \def \showarticletitle #1{#1}   \fi
\ifx \showURL      \undefined \def \showURL       {\relax}        \fi
% invisible to TeX
\providecommand\bibfield[2]{#2}
\providecommand\bibinfo[2]{#2}
\providecommand\natexlab[1]{#1}
\providecommand\showeprint[2][]{arXiv:#2}

\bibitem[\protect\citeauthoryear{Afonso and Prada}{Afonso and Prada}{2008}]%
        {afonso2008agents}
\bibfield{author}{\bibinfo{person}{Nuno Afonso} {and} \bibinfo{person}{Rui
  Prada}.} \bibinfo{year}{2008}\natexlab{}.
\newblock \showarticletitle{Agents that relate: Improving the social
  believability of non-player characters in role-playing games}. In
  \bibinfo{booktitle}{\emph{International Conference on Entertainment
  Computing}}. \bibinfo{publisher}{Springer}, \bibinfo{address}{Berlin,
  Heidelberg}, \bibinfo{pages}{34--45}.
\newblock


\bibitem[\protect\citeauthoryear{Ahmad, Bryant, Kleinman, Teng, Nguyen, and
  Seif El-Nasr}{Ahmad et~al\mbox{.}}{2019}]%
        {ahmad2019modeling}
\bibfield{author}{\bibinfo{person}{Sabbir Ahmad}, \bibinfo{person}{Andy
  Bryant}, \bibinfo{person}{Erica Kleinman}, \bibinfo{person}{Zhaoqing Teng},
  \bibinfo{person}{Truong-Huy~D. Nguyen}, {and} \bibinfo{person}{Magy Seif
  El-Nasr}.} \bibinfo{year}{2019}\natexlab{}.
\newblock \showarticletitle{Modeling Individual and Team Behavior through
  Spatio-Temporal Analysis}. In \bibinfo{booktitle}{\emph{Proceedings of the
  Annual Symposium on Computer-Human Interaction in Play}} (Barcelona, Spain)
  \emph{(\bibinfo{series}{CHI PLAY '19})}. \bibinfo{publisher}{Association for
  Computing Machinery}, \bibinfo{address}{New York, NY, USA},
  \bibinfo{pages}{601–612}.
\newblock
\showISBNx{9781450366885}
\urldef\tempurl%
\url{https://doi.org/10.1145/3311350.3347188}
\showDOI{\tempurl}


\bibitem[\protect\citeauthoryear{Baker, Saxe, and Tenenbaum}{Baker
  et~al\mbox{.}}{2011}]%
        {baker2011bayesian}
\bibfield{author}{\bibinfo{person}{Chris Baker}, \bibinfo{person}{Rebecca
  Saxe}, {and} \bibinfo{person}{Joshua Tenenbaum}.}
  \bibinfo{year}{2011}\natexlab{}.
\newblock \showarticletitle{Bayesian theory of mind: Modeling joint
  belief-desire attribution}. In \bibinfo{booktitle}{\emph{Proceedings of the
  annual meeting of the cognitive science society}}, Vol.~\bibinfo{volume}{33}.
\newblock
\newblock
\shownote{Retrieved from https://escholarship.org/uc/item/5rk7z59q.}


\bibitem[\protect\citeauthoryear{Baker}{Baker}{2012}]%
        {baker2012bayesian}
\bibfield{author}{\bibinfo{person}{Chris~Lawrence Baker}.}
  \bibinfo{year}{2012}\natexlab{}.
\newblock \emph{\bibinfo{title}{Bayesian theory of mind: Modeling human
  reasoning about beliefs, desires, goals, and social relations}}.
\newblock \bibinfo{thesistype}{Ph.D. Dissertation}.
  \bibinfo{school}{Massachusetts Institute of Technology}.
\newblock


\bibitem[\protect\citeauthoryear{Baker, Saxe, and Tenenbaum}{Baker
  et~al\mbox{.}}{2009}]%
        {baker2009action}
\bibfield{author}{\bibinfo{person}{Chris~L Baker}, \bibinfo{person}{Rebecca
  Saxe}, {and} \bibinfo{person}{Joshua~B Tenenbaum}.}
  \bibinfo{year}{2009}\natexlab{}.
\newblock \showarticletitle{Action understanding as inverse planning}.
\newblock \bibinfo{journal}{\emph{Cognition}} \bibinfo{volume}{113},
  \bibinfo{number}{3} (\bibinfo{year}{2009}), \bibinfo{pages}{329--349}.
\newblock


\bibitem[\protect\citeauthoryear{Baker and Tenenbaum}{Baker and
  Tenenbaum}{2014}]%
        {baker2014modeling}
\bibfield{author}{\bibinfo{person}{Chris~L Baker} {and}
  \bibinfo{person}{Joshua~B Tenenbaum}.} \bibinfo{year}{2014}\natexlab{}.
\newblock \showarticletitle{Modeling human plan recognition using Bayesian
  theory of mind}.
\newblock \bibinfo{journal}{\emph{Plan, activity, and intent recognition:
  Theory and practice}}  \bibinfo{volume}{7} (\bibinfo{year}{2014}),
  \bibinfo{pages}{177--204}.
\newblock


\bibitem[\protect\citeauthoryear{Bakkes, Spronck, and van Lankveld}{Bakkes
  et~al\mbox{.}}{2012}]%
        {bakkes2012player}
\bibfield{author}{\bibinfo{person}{Sander~CJ Bakkes},
  \bibinfo{person}{Pieter~HM Spronck}, {and} \bibinfo{person}{Giel van
  Lankveld}.} \bibinfo{year}{2012}\natexlab{}.
\newblock \showarticletitle{Player behavioural modelling for video games}.
\newblock \bibinfo{journal}{\emph{Entertainment Computing}}
  \bibinfo{volume}{3}, \bibinfo{number}{3} (\bibinfo{year}{2012}),
  \bibinfo{pages}{71--79}.
\newblock


\bibitem[\protect\citeauthoryear{Bard, Foerster, Chandar, Burch, Lanctot, Song,
  Parisotto, Dumoulin, Moitra, Hughes, et~al\mbox{.}}{Bard
  et~al\mbox{.}}{2020}]%
        {bard2020hanabi}
\bibfield{author}{\bibinfo{person}{Nolan Bard}, \bibinfo{person}{Jakob~N
  Foerster}, \bibinfo{person}{Sarath Chandar}, \bibinfo{person}{Neil Burch},
  \bibinfo{person}{Marc Lanctot}, \bibinfo{person}{H~Francis Song},
  \bibinfo{person}{Emilio Parisotto}, \bibinfo{person}{Vincent Dumoulin},
  \bibinfo{person}{Subhodeep Moitra}, \bibinfo{person}{Edward Hughes},
  {et~al\mbox{.}}} \bibinfo{year}{2020}\natexlab{}.
\newblock \showarticletitle{The hanabi challenge: A new frontier for ai
  research}.
\newblock \bibinfo{journal}{\emph{Artificial Intelligence}}
  \bibinfo{volume}{280} (\bibinfo{year}{2020}), \bibinfo{pages}{103216}.
\newblock


\bibitem[\protect\citeauthoryear{Berner, Brockman, Chan, Cheung, D{\k{e}}biak,
  Dennison, Farhi, Fischer, Hashme, Hesse, et~al\mbox{.}}{Berner
  et~al\mbox{.}}{2019}]%
        {berner2019dota}
\bibfield{author}{\bibinfo{person}{Christopher Berner}, \bibinfo{person}{Greg
  Brockman}, \bibinfo{person}{Brooke Chan}, \bibinfo{person}{Vicki Cheung},
  \bibinfo{person}{Przemys{\l}aw D{\k{e}}biak}, \bibinfo{person}{Christy
  Dennison}, \bibinfo{person}{David Farhi}, \bibinfo{person}{Quirin Fischer},
  \bibinfo{person}{Shariq Hashme}, \bibinfo{person}{Chris Hesse},
  {et~al\mbox{.}}} \bibinfo{year}{2019}\natexlab{}.
\newblock \showarticletitle{Dota 2 with large scale deep reinforcement
  learning}.
\newblock \bibinfo{journal}{\emph{arXiv preprint arXiv:1912.06680}}
  (\bibinfo{year}{2019}).
\newblock


\bibitem[\protect\citeauthoryear{Brown, Collins, and Duguid}{Brown
  et~al\mbox{.}}{1989}]%
        {brown1989situated}
\bibfield{author}{\bibinfo{person}{John~Seely Brown}, \bibinfo{person}{Allan
  Collins}, {and} \bibinfo{person}{Paul Duguid}.}
  \bibinfo{year}{1989}\natexlab{}.
\newblock \showarticletitle{Situated cognition and the culture of learning}.
\newblock \bibinfo{journal}{\emph{Educational researcher}}
  \bibinfo{volume}{18}, \bibinfo{number}{1} (\bibinfo{year}{1989}),
  \bibinfo{pages}{32--42}.
\newblock


\bibitem[\protect\citeauthoryear{Dautenhahn}{Dautenhahn}{1998}]%
        {dautenhahn1998art}
\bibfield{author}{\bibinfo{person}{Kerstin Dautenhahn}.}
  \bibinfo{year}{1998}\natexlab{}.
\newblock \showarticletitle{The art of designing socially intelligent agents:
  Science, fiction, and the human in the loop}.
\newblock \bibinfo{journal}{\emph{Applied artificial intelligence}}
  \bibinfo{volume}{12}, \bibinfo{number}{7-8} (\bibinfo{year}{1998}),
  \bibinfo{pages}{573--617}.
\newblock


\bibitem[\protect\citeauthoryear{Dimitrova}{Dimitrova}{2003}]%
        {dimitrova2003style}
\bibfield{author}{\bibinfo{person}{Vania Dimitrova}.}
  \bibinfo{year}{2003}\natexlab{}.
\newblock \showarticletitle{STyLE-OLM: Interactive open learner modelling}.
\newblock \bibinfo{journal}{\emph{International Journal of Artificial
  Intelligence in Education}} \bibinfo{volume}{13}, \bibinfo{number}{1}
  (\bibinfo{year}{2003}), \bibinfo{pages}{35--78}.
\newblock


\bibitem[\protect\citeauthoryear{El-Nasr, Drachen, and Canossa}{El-Nasr
  et~al\mbox{.}}{2016}]%
        {el2016game}
\bibfield{author}{\bibinfo{person}{Magy~Seif El-Nasr}, \bibinfo{person}{Anders
  Drachen}, {and} \bibinfo{person}{Alessandro Canossa}.}
  \bibinfo{year}{2016}\natexlab{}.
\newblock \bibinfo{booktitle}{\emph{Game analytics}}.
\newblock \bibinfo{publisher}{Springer}.
\newblock


\bibitem[\protect\citeauthoryear{Fails and Olsen~Jr}{Fails and
  Olsen~Jr}{2003}]%
        {fails2003interactive}
\bibfield{author}{\bibinfo{person}{Jerry~Alan Fails} {and}
  \bibinfo{person}{Dan~R Olsen~Jr}.} \bibinfo{year}{2003}\natexlab{}.
\newblock \showarticletitle{Interactive machine learning}. In
  \bibinfo{booktitle}{\emph{Proceedings of the 8th international conference on
  Intelligent user interfaces}}. \bibinfo{pages}{39--45}.
\newblock


\bibitem[\protect\citeauthoryear{Freeman and Wohn}{Freeman and Wohn}{2017}]%
        {freeman2017esports}
\bibfield{author}{\bibinfo{person}{Guo Freeman} {and}
  \bibinfo{person}{Donghee~Yvette Wohn}.} \bibinfo{year}{2017}\natexlab{}.
\newblock \showarticletitle{eSports as an emerging research context at CHI:
  Diverse perspectives on definitions}. In
  \bibinfo{booktitle}{\emph{Proceedings of the 2017 CHI conference extended
  abstracts on human factors in computing systems}}.
  \bibinfo{pages}{1601--1608}.
\newblock


\bibitem[\protect\citeauthoryear{F{\"u}rnkranz}{F{\"u}rnkranz}{2007}]%
        {furnkranz2007recent}
\bibfield{author}{\bibinfo{person}{Johannes F{\"u}rnkranz}.}
  \bibinfo{year}{2007}\natexlab{}.
\newblock \showarticletitle{Recent advances in machine learning and game
  playing}.
\newblock \bibinfo{journal}{\emph{{\"O}GAI Journal}} \bibinfo{volume}{26},
  \bibinfo{number}{2} (\bibinfo{year}{2007}), \bibinfo{pages}{19--28}.
\newblock


\bibitem[\protect\citeauthoryear{Gray}{Gray}{2017}]%
        {gray2017game}
\bibfield{author}{\bibinfo{person}{Wayne~D Gray}.}
  \bibinfo{year}{2017}\natexlab{}.
\newblock \bibinfo{title}{Game-XP: Action games as experimental paradigms for
  cognitive science}.
\newblock
\newblock


\bibitem[\protect\citeauthoryear{Hazelrigg}{Hazelrigg}{1998}]%
        {hazelrigg1998framework}
\bibfield{author}{\bibinfo{person}{George~A Hazelrigg}.}
  \bibinfo{year}{1998}\natexlab{}.
\newblock \showarticletitle{A framework for decision-based engineering design}.
\newblock  (\bibinfo{year}{1998}).
\newblock


\bibitem[\protect\citeauthoryear{Hodge, Devlin, Sephton, Block, Drachen, and
  Cowling}{Hodge et~al\mbox{.}}{2017}]%
        {hodge2017win}
\bibfield{author}{\bibinfo{person}{Victoria Hodge}, \bibinfo{person}{Sam
  Devlin}, \bibinfo{person}{Nick Sephton}, \bibinfo{person}{Florian Block},
  \bibinfo{person}{Anders Drachen}, {and} \bibinfo{person}{Peter Cowling}.}
  \bibinfo{year}{2017}\natexlab{}.
\newblock \showarticletitle{Win prediction in esports: Mixed-rank match
  prediction in multi-player online battle arena games}.
\newblock \bibinfo{journal}{\emph{arXiv preprint arXiv:1711.06498}}
  (\bibinfo{year}{2017}).
\newblock


\bibitem[\protect\citeauthoryear{Hodge, Devlin, Sephton, Block, Cowling, and
  Drachen}{Hodge et~al\mbox{.}}{2019}]%
        {hodge2019win}
\bibfield{author}{\bibinfo{person}{Victoria~J Hodge},
  \bibinfo{person}{Sam~Michael Devlin}, \bibinfo{person}{Nicholas~John
  Sephton}, \bibinfo{person}{Florian~Oliver Block}, \bibinfo{person}{Peter~Ivan
  Cowling}, {and} \bibinfo{person}{Anders Drachen}.}
  \bibinfo{year}{2019}\natexlab{}.
\newblock \showarticletitle{Win prediction in multi-player esports: Live
  professional match prediction}.
\newblock \bibinfo{journal}{\emph{IEEE Transactions on Games}}
  (\bibinfo{year}{2019}).
\newblock


\bibitem[\protect\citeauthoryear{Holzinger}{Holzinger}{2016}]%
        {holzinger2016interactive}
\bibfield{author}{\bibinfo{person}{Andreas Holzinger}.}
  \bibinfo{year}{2016}\natexlab{}.
\newblock \showarticletitle{Interactive machine learning for health
  informatics: when do we need the human-in-the-loop?}
\newblock \bibinfo{journal}{\emph{Brain Informatics}} \bibinfo{volume}{3},
  \bibinfo{number}{2} (\bibinfo{year}{2016}), \bibinfo{pages}{119--131}.
\newblock


\bibitem[\protect\citeauthoryear{Jaderberg, Czarnecki, Dunning, Marris, Lever,
  Castaneda, Beattie, Rabinowitz, Morcos, Ruderman, et~al\mbox{.}}{Jaderberg
  et~al\mbox{.}}{2019}]%
        {jaderberg2019human}
\bibfield{author}{\bibinfo{person}{Max Jaderberg}, \bibinfo{person}{Wojciech~M
  Czarnecki}, \bibinfo{person}{Iain Dunning}, \bibinfo{person}{Luke Marris},
  \bibinfo{person}{Guy Lever}, \bibinfo{person}{Antonio~Garcia Castaneda},
  \bibinfo{person}{Charles Beattie}, \bibinfo{person}{Neil~C Rabinowitz},
  \bibinfo{person}{Ari~S Morcos}, \bibinfo{person}{Avraham Ruderman},
  {et~al\mbox{.}}} \bibinfo{year}{2019}\natexlab{}.
\newblock \showarticletitle{Human-level performance in 3D multiplayer games
  with population-based reinforcement learning}.
\newblock \bibinfo{journal}{\emph{Science}} \bibinfo{volume}{364},
  \bibinfo{number}{6443} (\bibinfo{year}{2019}), \bibinfo{pages}{859--865}.
\newblock


\bibitem[\protect\citeauthoryear{Larsson}{Larsson}{2003}]%
        {larsson2003making}
\bibfield{author}{\bibinfo{person}{Andreas Larsson}.}
  \bibinfo{year}{2003}\natexlab{}.
\newblock \showarticletitle{Making sense of collaboration: the challenge of
  thinking together in global design teams}. In
  \bibinfo{booktitle}{\emph{Proceedings of the 2003 international ACM SIGGROUP
  conference on Supporting group work}}. \bibinfo{pages}{153--160}.
\newblock


\bibitem[\protect\citeauthoryear{Machado, Fantini, and Chaimowicz}{Machado
  et~al\mbox{.}}{2011a}]%
        {machado2011player}
\bibfield{author}{\bibinfo{person}{Marlos~C Machado},
  \bibinfo{person}{Eduardo~PC Fantini}, {and} \bibinfo{person}{Luiz
  Chaimowicz}.} \bibinfo{year}{2011}\natexlab{a}.
\newblock \showarticletitle{Player modeling: Towards a common taxonomy}. In
  \bibinfo{booktitle}{\emph{2011 16th international conference on computer
  games (CGAMES)}}. IEEE, \bibinfo{pages}{50--57}.
\newblock


\bibitem[\protect\citeauthoryear{Machado, Fantini, and Chaimowicz}{Machado
  et~al\mbox{.}}{2011b}]%
        {machado2011player2}
\bibfield{author}{\bibinfo{person}{Marlos~C Machado},
  \bibinfo{person}{Eduardo~PC Fantini}, {and} \bibinfo{person}{Luiz
  Chaimowicz}.} \bibinfo{year}{2011}\natexlab{b}.
\newblock \showarticletitle{Player Modeling: What is it? How to do it?}
\newblock \bibinfo{journal}{\emph{Proceedings of SBGames}}
  (\bibinfo{year}{2011}).
\newblock


\bibitem[\protect\citeauthoryear{Millis, Forsyth, Wallace, Graesser, and
  Timmins}{Millis et~al\mbox{.}}{2017}]%
        {millis2017impact}
\bibfield{author}{\bibinfo{person}{Keith Millis}, \bibinfo{person}{Carol
  Forsyth}, \bibinfo{person}{Patricia Wallace}, \bibinfo{person}{Arthur~C
  Graesser}, {and} \bibinfo{person}{Gary Timmins}.}
  \bibinfo{year}{2017}\natexlab{}.
\newblock \showarticletitle{The impact of game-like features on learning from
  an intelligent tutoring system}.
\newblock \bibinfo{journal}{\emph{Technology, Knowledge and Learning}}
  \bibinfo{volume}{22}, \bibinfo{number}{1} (\bibinfo{year}{2017}),
  \bibinfo{pages}{1--22}.
\newblock


\bibitem[\protect\citeauthoryear{Nguyen, Subramanian, El-Nasr, and
  Canossa}{Nguyen et~al\mbox{.}}{2014}]%
        {nguyen2014strategy}
\bibfield{author}{\bibinfo{person}{Truong-Huy~D Nguyen}, \bibinfo{person}{Shree
  Subramanian}, \bibinfo{person}{Magy~Seif El-Nasr}, {and}
  \bibinfo{person}{Alessandro Canossa}.} \bibinfo{year}{2014}\natexlab{}.
\newblock \showarticletitle{Strategy Detection in Wuzzit: A Decision Theoretic
  Approach}. In \bibinfo{booktitle}{\emph{International Conference on Learning
  ScienceWorkshop on Learning Analytics for Learning and Becoming a Practice}}.
\newblock


\bibitem[\protect\citeauthoryear{Nunes, Zhang, and Silva}{Nunes
  et~al\mbox{.}}{2015}]%
        {nunes2015survey}
\bibfield{author}{\bibinfo{person}{David~Sousa Nunes}, \bibinfo{person}{Pei
  Zhang}, {and} \bibinfo{person}{Jorge~S{\'a} Silva}.}
  \bibinfo{year}{2015}\natexlab{}.
\newblock \showarticletitle{A survey on human-in-the-loop applications towards
  an internet of all}.
\newblock \bibinfo{journal}{\emph{IEEE Communications Surveys \& Tutorials}}
  \bibinfo{volume}{17}, \bibinfo{number}{2} (\bibinfo{year}{2015}),
  \bibinfo{pages}{944--965}.
\newblock


\bibitem[\protect\citeauthoryear{{Omnicoach.gg}}{{Omnicoach.gg}}{2021}]%
        {omnicoach}
\bibfield{author}{\bibinfo{person}{{Omnicoach.gg}}.}
  \bibinfo{year}{2021}\natexlab{}.
\newblock \bibinfo{title}{Omnicoach.GG - AI Coach with Video Guides for
  Gaming}.
\newblock \bibinfo{howpublished}{\url{https://www.omnicoach.gg/}}.
\newblock
\newblock
\shownote{[Online; accessed 12-Feb-2021].}


\bibitem[\protect\citeauthoryear{Pedraza-Ramirez, Musculus, Raab, and
  Laborde}{Pedraza-Ramirez et~al\mbox{.}}{2020}]%
        {pedraza2020setting}
\bibfield{author}{\bibinfo{person}{Ismael Pedraza-Ramirez},
  \bibinfo{person}{Lisa Musculus}, \bibinfo{person}{Markus Raab}, {and}
  \bibinfo{person}{Sylvain Laborde}.} \bibinfo{year}{2020}\natexlab{}.
\newblock \showarticletitle{Setting the scientific stage for esports
  psychology: a systematic review}.
\newblock \bibinfo{journal}{\emph{International Review of Sport and Exercise
  Psychology}} \bibinfo{volume}{13}, \bibinfo{number}{1}
  (\bibinfo{year}{2020}), \bibinfo{pages}{319--352}.
\newblock


\bibitem[\protect\citeauthoryear{Pluss, Bennett, Novak, Panchuk, Coutts, and
  Fransen}{Pluss et~al\mbox{.}}{2019}]%
        {pluss2019esports}
\bibfield{author}{\bibinfo{person}{Matthew~A Pluss}, \bibinfo{person}{Kyle~JM
  Bennett}, \bibinfo{person}{Andrew~R Novak}, \bibinfo{person}{Derek Panchuk},
  \bibinfo{person}{Aaron~J Coutts}, {and} \bibinfo{person}{Job Fransen}.}
  \bibinfo{year}{2019}\natexlab{}.
\newblock \showarticletitle{Esports: the chess of the 21st century}.
\newblock \bibinfo{journal}{\emph{Frontiers in psychology}}
  \bibinfo{volume}{10} (\bibinfo{year}{2019}), \bibinfo{pages}{156}.
\newblock


\bibitem[\protect\citeauthoryear{Premack and Woodruff}{Premack and
  Woodruff}{1978}]%
        {premack1978does}
\bibfield{author}{\bibinfo{person}{David Premack} {and} \bibinfo{person}{Guy
  Woodruff}.} \bibinfo{year}{1978}\natexlab{}.
\newblock \showarticletitle{Does the chimpanzee have a theory of mind?}
\newblock \bibinfo{journal}{\emph{Behavioral and brain sciences}}
  \bibinfo{volume}{1}, \bibinfo{number}{4} (\bibinfo{year}{1978}),
  \bibinfo{pages}{515--526}.
\newblock


\bibitem[\protect\citeauthoryear{Risi and Preuss}{Risi and Preuss}{2020}]%
        {risi2020chess}
\bibfield{author}{\bibinfo{person}{Sebastian Risi} {and} \bibinfo{person}{Mike
  Preuss}.} \bibinfo{year}{2020}\natexlab{}.
\newblock \showarticletitle{From chess and atari to starcraft and beyond: How
  game AI is driving the world of AI}.
\newblock \bibinfo{journal}{\emph{KI-K{\"u}nstliche Intelligenz}}
  \bibinfo{volume}{34}, \bibinfo{number}{1} (\bibinfo{year}{2020}),
  \bibinfo{pages}{7--17}.
\newblock


\bibitem[\protect\citeauthoryear{{Senpai.gg}}{{Senpai.gg}}{2021}]%
        {senpai}
\bibfield{author}{\bibinfo{person}{{Senpai.gg}}.}
  \bibinfo{year}{2021}\natexlab{}.
\newblock \bibinfo{title}{SenpAI.GG - AI Coach with Video Guides for Gaming}.
\newblock \bibinfo{howpublished}{\url{https://www.senpai.gg/}}.
\newblock
\newblock
\shownote{[Online; accessed 12-Feb-2021].}


\bibitem[\protect\citeauthoryear{Shergadwala, Bilionis, Kannan, and
  Panchal}{Shergadwala et~al\mbox{.}}{2018b}]%
        {shergadwala2018quantifying}
\bibfield{author}{\bibinfo{person}{Murtuza Shergadwala}, \bibinfo{person}{Ilias
  Bilionis}, \bibinfo{person}{Karthik~N Kannan}, {and}
  \bibinfo{person}{Jitesh~H Panchal}.} \bibinfo{year}{2018}\natexlab{b}.
\newblock \showarticletitle{Quantifying the impact of domain knowledge and
  problem framing on sequential decisions in engineering design}.
\newblock \bibinfo{journal}{\emph{Journal of Mechanical Design}}
  \bibinfo{volume}{140}, \bibinfo{number}{10} (\bibinfo{year}{2018}).
\newblock


\bibitem[\protect\citeauthoryear{Shergadwala, Bilionis, and
  Panchal}{Shergadwala et~al\mbox{.}}{2018a}]%
        {shergadwala2018students}
\bibfield{author}{\bibinfo{person}{Murtuza Shergadwala}, \bibinfo{person}{Ilias
  Bilionis}, {and} \bibinfo{person}{Jitesh~H Panchal}.}
  \bibinfo{year}{2018}\natexlab{a}.
\newblock \showarticletitle{Students as sequential decision-makers: Quantifying
  the impact of problem knowledge and process deviation on the achievement of
  their design problem objective}. In \bibinfo{booktitle}{\emph{International
  Design Engineering Technical Conferences and Computers and Information in
  Engineering Conference}}, Vol.~\bibinfo{volume}{51784}. American Society of
  Mechanical Engineers, \bibinfo{pages}{V003T04A011}.
\newblock


\bibitem[\protect\citeauthoryear{Shergadwala, Kannan, and Panchal}{Shergadwala
  et~al\mbox{.}}{[n.d.]}]%
        {shergadwala2016understanding}
\bibfield{author}{\bibinfo{person}{Murtuza Shergadwala},
  \bibinfo{person}{Karthik~N Kannan}, {and} \bibinfo{person}{Jitesh~H
  Panchal}.} \bibinfo{year}{[n.d.]}\natexlab{}.
\newblock \showarticletitle{Understanding the impact of expertise on design
  outcome: An approach based on concept inventories and item response theory}.
  In \bibinfo{booktitle}{\emph{ASME 2016 International Design Engineering
  Technical Conferences and Computers and Information in Engineering
  Conference}}. American Society of Mechanical Engineers Digital Collection.
\newblock


\bibitem[\protect\citeauthoryear{Simon}{Simon}{1988}]%
        {simon1988science}
\bibfield{author}{\bibinfo{person}{Herbert~A Simon}.}
  \bibinfo{year}{1988}\natexlab{}.
\newblock \showarticletitle{The science of design: Creating the artificial}.
\newblock \bibinfo{journal}{\emph{Design Issues}} (\bibinfo{year}{1988}),
  \bibinfo{pages}{67--82}.
\newblock


\bibitem[\protect\citeauthoryear{Sukthankar, Geib, Bui, Pynadath, and
  Goldman}{Sukthankar et~al\mbox{.}}{2014}]%
        {sukthankar2014plan}
\bibfield{author}{\bibinfo{person}{Gita Sukthankar},
  \bibinfo{person}{Christopher Geib}, \bibinfo{person}{Hung~Hai Bui},
  \bibinfo{person}{David Pynadath}, {and} \bibinfo{person}{Robert~P Goldman}.}
  \bibinfo{year}{2014}\natexlab{}.
\newblock \bibinfo{booktitle}{\emph{Plan, activity, and intent recognition:
  Theory and practice}}.
\newblock \bibinfo{publisher}{Newnes}.
\newblock


\bibitem[\protect\citeauthoryear{van~den Herik, Donkers, and Spronck}{van~den
  Herik et~al\mbox{.}}{2005}]%
        {van2005opponent}
\bibfield{author}{\bibinfo{person}{H~Jaap van~den Herik}, \bibinfo{person}{HHLM
  Donkers}, {and} \bibinfo{person}{Pieter~HM Spronck}.}
  \bibinfo{year}{2005}\natexlab{}.
\newblock \showarticletitle{Opponent modelling and commercial games}.
\newblock \bibinfo{journal}{\emph{Proc. IEEE}} (\bibinfo{year}{2005}),
  \bibinfo{pages}{15--25}.
\newblock


\bibitem[\protect\citeauthoryear{Wang, Churchill, Maes, Fan, Shneiderman, Shi,
  and Wang}{Wang et~al\mbox{.}}{2020}]%
        {wang2020human}
\bibfield{author}{\bibinfo{person}{Dakuo Wang}, \bibinfo{person}{Elizabeth
  Churchill}, \bibinfo{person}{Pattie Maes}, \bibinfo{person}{Xiangmin Fan},
  \bibinfo{person}{Ben Shneiderman}, \bibinfo{person}{Yuanchun Shi}, {and}
  \bibinfo{person}{Qianying Wang}.} \bibinfo{year}{2020}\natexlab{}.
\newblock \showarticletitle{From Human-Human Collaboration to Human-AI
  Collaboration: Designing AI Systems That Can Work Together with People}. In
  \bibinfo{booktitle}{\emph{Extended Abstracts of the 2020 CHI Conference on
  Human Factors in Computing Systems}}. \bibinfo{pages}{1--6}.
\newblock


\bibitem[\protect\citeauthoryear{Wang}{Wang}{2018}]%
        {wang2018predictive}
\bibfield{author}{\bibinfo{person}{Tian Wang}.}
  \bibinfo{year}{2018}\natexlab{}.
\newblock \showarticletitle{Predictive Analysis on eSports Games: A Case Study
  on League of Legends (LoL) eSports Tournaments}.
\newblock  (\bibinfo{year}{2018}).
\newblock


\bibitem[\protect\citeauthoryear{Weiss and Schiele}{Weiss and Schiele}{2013}]%
        {weiss2013virtual}
\bibfield{author}{\bibinfo{person}{Thomas Weiss} {and} \bibinfo{person}{Sabrina
  Schiele}.} \bibinfo{year}{2013}\natexlab{}.
\newblock \showarticletitle{Virtual worlds in competitive contexts: Analyzing
  eSports consumer needs}.
\newblock \bibinfo{journal}{\emph{Electronic Markets}} \bibinfo{volume}{23},
  \bibinfo{number}{4} (\bibinfo{year}{2013}), \bibinfo{pages}{307--316}.
\newblock


\bibitem[\protect\citeauthoryear{Yannakakis, Spronck, Loiacono, and
  Andr{\'e}}{Yannakakis et~al\mbox{.}}{2013}]%
        {yannakakis2013player}
\bibfield{author}{\bibinfo{person}{Georgios~N Yannakakis},
  \bibinfo{person}{Pieter Spronck}, \bibinfo{person}{Daniele Loiacono}, {and}
  \bibinfo{person}{Elisabeth Andr{\'e}}.} \bibinfo{year}{2013}\natexlab{}.
\newblock \showarticletitle{Player modeling}.
\newblock  (\bibinfo{year}{2013}).
\newblock


\bibitem[\protect\citeauthoryear{Yannakakis and Togelius}{Yannakakis and
  Togelius}{2015}]%
        {yannakakis2015experience}
\bibfield{author}{\bibinfo{person}{Georgios~N Yannakakis} {and}
  \bibinfo{person}{Julian Togelius}.} \bibinfo{year}{2015}\natexlab{}.
\newblock \showarticletitle{Experience-driven procedural content generation}.
  In \bibinfo{booktitle}{\emph{2015 International Conference on Affective
  Computing and Intelligent Interaction (ACII)}}. IEEE,
  \bibinfo{pages}{519--525}.
\newblock


\bibitem[\protect\citeauthoryear{Yoshida, Dolan, and Friston}{Yoshida
  et~al\mbox{.}}{2008}]%
        {yoshida2008game}
\bibfield{author}{\bibinfo{person}{Wako Yoshida}, \bibinfo{person}{Ray~J
  Dolan}, {and} \bibinfo{person}{Karl~J Friston}.}
  \bibinfo{year}{2008}\natexlab{}.
\newblock \showarticletitle{Game theory of mind}.
\newblock \bibinfo{journal}{\emph{PLoS Comput Biol}} \bibinfo{volume}{4},
  \bibinfo{number}{12} (\bibinfo{year}{2008}), \bibinfo{pages}{e1000254}.
\newblock


\end{thebibliography}
\end{document}